\pdfoutput=1
\documentclass[useAMS,usenatbib,usegraphicx]{mn2e}
\usepackage{amssymb}
\usepackage{amsmath}
\usepackage{ctable}
\usepackage{fixltx2e} 
\usepackage{bm}

\usepackage{graphicx}	
\usepackage{mathrsfs}

\newcommand{\pd}[2]{\ensuremath{\partial #1/\partial #2}} 
\newcommand{\pdif}[2]{\ensuremath{ \frac{\partial #1}{\partial #2}}} 
\newcommand{\f}[2]{\frac{#1}{#2}} 
\newcommand{\vv}[1]{\ensuremath{\mathbf{#1}}}
\newcommand{\vor}{\bm\omega}
\newcommand{\beq}{\begin{equation}}
\newcommand{\seq}{\end{equation}}

\newcommand{\cv}{c_{\rm{v}}}

\newcommand{\athena}{\textsc{Athena{\scriptsize ++ }}}

\title[Outflows from Inflows]
{Outflows from inflows: the nature of Bondi-like accretion}
\author[Waters et al.]
{
   Tim Waters$^1$\thanks{E-mail: waters@lanl.gov},
   Aycin Aykutalp$^1$,
   Daniel Proga$^2$,
   Jarrett Johnson$^1$,
   Hui Li$^1$,
   \newauthor
    \hspace{2pt} and Joseph Smidt$^1$
  \\
$^1$Center for Theoretical Astrophysics, Los Alamos National Laboratory, Los Alamos, NM, 87545, USA\\
$^2$Department of Physics \& Astronomy, University of Nevada, Las Vegas
4505 S. Maryland Pkwy, Las Vegas, NV, 89154-4002, USA\\
}

\date{Submitted 2 October 2019; accepted \today}

\begin{document}
\maketitle
\label{firstpage}

\begin{abstract}
The classic Bondi solution remains a common starting point both for studying black hole growth across cosmic time in cosmological simulations and for smaller scale simulations of AGN feedback.
In nature, however, there will be inhomogenous distributions of rotational velocity and density along the outer radius ($R_o$) marking the sphere of influence of a black hole.  While there have been many studies of how the Bondi solution changes with a prescribed angular momentum boundary condition, they have all assumed a constant density at $R_o$.  
In this Letter, we show that a non-uniform density at $R_o$ causes a meridional flow and due to conservation of angular momentum, the Bondi solution qualitatively changes into an inflow-outflow solution.  
Using physical arguments, we analytically identify the critical logarithmic density gradient $|\pd{\ln{\rho}}{\theta}|$ above which this change of the solution occurs.  For realistic $R_o$, this critical gradient is less than 0.01 and tends to 0 as $R_o \rightarrow \infty$. 
We show using numerical simulations that, unlike for solutions with an imposed rotational velocity, the accretion rate for solutions under an inhomogenous density boundary condition remains constant at nearly the Bondi rate $\dot{M}_B$, while the outflow rate can greatly exceed $\dot{M}_B$.  

\end{abstract}


\section{Introduction}
There are multiple modes of accretion thought to be important for black hole (BH) growth.  
The so-called hot accretion mode refers to disk mediated accretion through geometrically thick disks that are either optically thin (as in ADAFs; Narayan \& Yi 1999; see Yuan \& Narayan 2014 for a review) or optically thick (as in slim disks; Abramowicz et al. 1988).  Likewise, the cold accretion mode refers to standard geometrically thin, optically thick disks (Shakura \& Sunyaev 1971). 
Bondi accretion (Bondi 1952), meanwhile, implies very low angular momentum accretion --- low enough that the gas encounters no angular momentum barrier on its way to the BH.  
Realistically, spherically symmetric Bondi accretion never occurs in nature, but three-dimensional (3D) time-dependent numerical simulations reveal that Bondi accretion can occur along the rotation axis of hot mode accretion flows (Janiuk, Proga, \& Kurosawa 2008).  

Accretion disks themselves are thought to form from `Bondi-like' accretion, in which a flow with low angular momentum accretes approximately spherically within the Bondi radius until it circularizes upon encountering a centrifugal barrier.  
Proga \& Begelman (2003a) first explored this numerically by solving a specific instance of the `inhomogenous Bondi problem', 
adopting simple prescriptions for the azimuthal velocity, $v_\phi(r=R_o,\theta)$, i.e. for the $\theta$-component of angular momentum ($l_\theta$).
Their simulations confirmed the intuitive analytic results of Abramowicz \& Zurek (1981): accretion at the Bondi rate can only occur if the gas does not encounter a centrifugal barrier, while hot mode accretion with a greatly reduced accretion rate results otherwise (unless the effective viscosity is quite large; see Narayan \& Fabian 2011).  
Subsequent works have considered alternative prescriptions for angular momentum (e.g., Krumholz, McKee \& Klein 2005), magnetohydrodynamic (MHD) evolution (Proga \& Begelman 2003b; Igumenshchev, Narayan \& Abramowicz 2003), and general relativistic effects for spinning BHs (Palit, Janiuk \& Sukova 2019).  In the last decade there have been numerous studies of low angular momentum accretion that include radiative feedback processes (e.g., Kurosowa \& Proga 2009; Ciotti et al. 2009, 2010; Park \& Ricotti 2011; Li et al. 2013; Ciotti et al. 2017; Yuan et al. 2018; Bu \& Yang 2019a,b,c; Yoon et al. 2019; Gan et al. 2019), and this Bondi-like modeling framework has shown promise for reproducing the observed X-ray luminosity across many orders of magnitude in BH mass in early type galaxies (e.g., Li et al. 2018; Pellegrini et al. 2019).

The present study points to a need to reconsider the notion that accretion disks themselves can form via Bondi-like accretion.
In general, the inhomogenous Bondi problem has boundary conditions (BCs) defined on a spherical surface with a nonzero surface gradient (or surface curl) of density or pressure in addition to velocity.  
In this letter, we investigate the other `simple' axisymmetric problem: we consider a density BC with $\pd{\rho}{\theta} \neq 0$ rather than breaking spherical symmetry by considering $\pd{v_\phi}{\theta} \neq 0$ at $R_o$.   
A non-uniform density distribution at $R_o$ gives rise to pressure gradients in the flow that result in a nonzero $v_\theta$.
The meridional flows that develop are not at all analogous to the azimuthal flows in simulations with small nonzero $l_\theta$.  The streamlines have the opposite curvature, meaning an inflowing streamline turns into an outflowing one rather than circling the BH.  Inflow-outflow solutions mediated by disks have long been speculated to occur due to the fact that the Bernoulli function is positive (Blandford \& Begelman 1999, 2004; Begelman 2012), but here we are simply dealing with internal angular momentum transport without a disk. 

This letter is organized as follows.  In \S{2} we show explicitly that the inhomogenous Bondi problem can be considered an angular momentum transport problem, and we derive the critical density gradient that will result in inflow-outflow solutions.  In \S{3} we present our simulation results, and 
in \S{4} we discuss the overall flow properties qualitatively.  

\section{The inhomogenous Bondi problem} 
The role of angular momentum in this problem is perhaps best appreciated through a quote
from R. Kulsrud's textbook (Kulsrud 2005).  Referring to inviscid flows, he states that ``steady state accretion cannot occur unless there is a magnetic torque that removes angular momentum."  
First defining the total angular momentum in a general volume $V$ by $\vv{H} = \int_V \rho \, \bm{\ell}\, dV$, where $\bm{\ell}= \vv{r}\times\vv{v}$ is the specific angular momentum of a fluid element located a distance $\vv{r}$ away from some origin $O$, Kulsrud derives an equation for $d\vv{H}/dt$ starting from the momentum equation in ideal MHD
\begin{equation}
\pdif{\left(\rho\mathbf{v}\right)}{t} + \mathbf{\nabla} \cdot \left( \rho\mathbf{v} 
\mathbf{v} + p\,\mathbb{I} + \vv{T}_B\right) = \rho \vv{g},\label{eq:force}
\end{equation}
where $p$ is the gas pressure, $\vv{T}_B =B^2/8\pi  \mathbb{I} - \vv{B}\vv{B}/4\pi$ is the magnetic stress tensor (with $\mathbb{I}$ the unit tensor and $\vv{B}$ the magnetic field), and $\vv{g}$ is the gravitational force.  Namely, substituting for $\partial \rho\vv{v}/\partial t$ in the expression for $d\vv{H}/dt$ using  eq. \eqref{eq:force}, Kulsrud arrives at 
\begin{equation}
 \f{d\mathbf{H}}{dt} = -\int_S \rho\, \bm{\ell}\, \vv{v}\cdot d\vv{S} + \int_S \vv{r}\times\vv{B} \f{\vv{B}}{4\pi} \cdot d\vv{S}.
 \label{dHdt}
 \end{equation}
Here, $S$ is a spherical surface enclosing $V$.  The only assumption used to derive eq. \eqref{dHdt} is that gravity is a central force, $\vv{g} = -\nabla \Phi$. 
Hence, we see that in the absence of any viscosity, steady state accretion ($d\vv{H}/dt = 0$) can only be achieved if the above integrals are equal.  This is the logic underlying the above quote.  However, spherical, unmagnetized accretion (i.e. the Bondi solution) is the very special case $\vv{B} = 0$ and $\bm{\ell} = 0$, meaning the equality of eq. \eqref{dHdt} is trivially satisfied.  
The only exception to Kulsrud's statement involves unmagnetized cases in which the integral  
$\int_S \rho\, \bm{\ell}\, \vv{v}\cdot d\vv{S}$ vanishes due to internal dynamics in flows with nonzero specific angular momentum.  Such is the nature of steady state solutions to the inhomogenous Bondi problem; every spherical surface in $V$ must satisfy $\int_S \rho\, \bm{\ell}\, \vv{v}\cdot d\vv{S} =0$.  

\subsection{The critical density gradient}
A BC with $\pd{\rho}{\theta} \neq 0$ at $R_o$ leads to $l_\phi \neq 0$, and we therefore seek steady state solutions with $v_\phi = 0$ but $v_\theta \neq 0$ at $R_o$.  
To identify a consistent set of BCs, in the appendix we derive the basic equations governing steady state solutions.  In particular, eq. \eqref{lphi_phi} governs arbitrary axisymmetric BCs, and for a homentropic BC ($\pd{s}{\theta} = 0$) with $v_\phi = 0$
it reads
\beq
l_\phi = -R_o\f{\pd{h}{\theta} + R_o\, v_r \, \pd{v_\theta}{r} }{v_r  + \pd{v_\theta}{\theta}} .
\label{lphi_s}
\seq
For solutions obeying a standard outflow BC $\pd{v_\theta}{r} = 0$, we can place a bound on the maximum density gradient that will result in pure inflow.\footnote{The enthalpy of an ideal gas is $h = c_s^2/(\gamma - 1) = \gamma p \rho^{-\gamma} \rho^{\gamma-1}/(\gamma - 1)$, with $c_s^2 = \gamma p/\rho$.  Thus, taking the partial derivative $\pd{h}{\theta}$ at constant entropy gives  $\pd{h}{\theta} = c_s^2 \pd{\ln\rho}{\theta}$.}  
To accrete from large distances directly onto the BH, gas must have a specific angular momentum $|\bm{\ell}| \lesssim 2 R_s\,c$ (Abramowicz \& Zurek 1981).  Enforcing this bound using eq. \eqref{lphi_s}, we find
\beq
\left| \pdif{\ln\rho_{\rm{bdy}}}{\theta} \right| \lesssim \f{4R_B}{R_o}   \f{T_0}{T(R_o,\theta)}  \left| \f{v_r  + \pd{v_\theta}{\theta}}{c} \right|_{r = R_o} ,
\label{rho_bound}
\seq
where $R_B = GM_{\rm{bh}}/c_{s,0}^2$ is the Bondi radius for a BH with mass $M_{\rm{bh}}$ (with $c_{s,0} = (\gamma p_0/\rho_0)^{1/2}$ the adiabatic sound speed for a reference temperature $T_0 = k_B^{-1} \bar{m} p_0/\rho_0$ at $R_0$).  
We have arrived at our main result: the quantity on the right hand side is in general a tiny number.  In the limiting case $R_o \rightarrow \infty$ and $v_r(R_o) \rightarrow 0$ considered by Bondi (1952), $|\pd{\ln\rho_{\rm{bdy}}}{\theta}|$ can be infinitesimally small and yet still qualitatively change the nature of the Bondi solution by turning part of the inflow into an outflow, as we now demonstrate. 

\vspace{-0.1in}
\section{Simulations}
Using \athena (Release 19.0; Stone et al. 2019, in prep.), we solve the equations of adiabatic hydrodynamics, i.e. the force equation given in eq. \eqref{eq:force} with $\vv{B} = 0$, the continuity equation $\pd{\rho}{t} + \nabla \cdot \rho \vv{v} = 0$, and the entropy equation $\pd{s}{t} + \vv{v}\cdot \nabla s = 0$, where $s = c_{\rm{v}} \ln p/\rho^\gamma$ is the specific entropy (with $\cv$ the specific heat at constant volume and $\gamma$ the adiabatic index).  We examine solutions with $\gamma = 5/3$, implementing the Paczy{\'n}sky \& Wiita pseudo-Newtonian potential, $\phi = -GM_{\rm{bh}}/(r-R_s)$ (Paczy{\'n}sky \& Wiita 1980), to place the sonic point outside the Schwarzschild radius, $R_s = 2GM_{\rm{bh}}/c^2$. 
We perform 2D axisymmetric calculations in spherical polar coordinates $(r,\theta,\phi)$ on a uniformly spaced polar grid with $ N_\theta = 270$ zones from $0$ to $\pi/2$ and a logarithmic radial grid with $N_r = 400$ zones and $dr_{i+1}/dr_i = 1.019$ from $r_{\rm{in}} = 10R_s$ to $r_{\rm{out}} = R_o$. 
We apply reflecting BCs at $\theta = 0,\pi/2$ and `constant gradient BCs' ($\pd{q}{r} = constant$) in all primitive variables at both the inner and outer radius unless otherwise noted.  Finally, we use the 3rd order time integration scheme and the HLLC Riemann solver.

Solutions to the classic Bondi problem depend only on $\gamma$ and $R_B$, whereas solutions with a boundary at finite $R_o$ also depend on the value of $R_o/R_B$.  This dependence has been recently explored by Samadi et al. (2019) and the solution procedure has been detailed by Waters \& Proga (2012) in the context of Parker winds.  
Our initial conditions (ICs) are $v_\theta = v_\phi = 0$, while $v_r$, $\rho$, and $p$ are the semi-analytic solutions to this generalized 1D Bondi problem under a BC $\rho_{\rm{bdy}}(R_0,\theta)$ given below; the code for these ICs is publicly available at github.com/trwaters/bondiparker.  
We set $R_o = 8 R_B$, noting that the outflow rates we find will have a dependence on this parameter since gas becomes less bound at larger radii (see \S{4}). 

We allow for an arbitrary density distribution to be specified along $R_o$ using the function
\beq \rho_{\rm{bdy}}(\theta) = \rho_0\, \mathcal{N} f(\theta) , \seq 
where $f(\theta)$ is a distribution function and $\mathcal{N}^{-1} = (1/2) \int_0^\pi f(\theta) \sin\theta d\theta$ is a normalization integral.  The function $f$ is normalized to recover the classic Bondi solution when $f(\theta) = 1$, i.e. $\int_0^\pi f(\theta)\,d\theta = \pi$.  The normalization integral is chosen to satisfy $M_R = \int \rho_{\rm{bdy}}(\theta)\,dV$ 
($\approx 2\pi \rho_0 \mathcal{N} \Delta r R_o^2 \int_0^\pi f(\theta) \sin\theta d\theta$ for $\Delta r \ll R_o$), 
where $M_R = 4\pi R_o^2 \Delta r \,\rho_0$ is the mass of a gas reservoir (a shell of matter extending from $R_o$ to $R_o + \Delta r$ that is constantly replinished) that is implicitly assumed to exist when applying this BC.  By requiring that the mass of the reservoir is the same across all density distributions, we can properly compare the resulting accretion and outflow rates with the Bondi rate.

We consider the bell-shaped function $f(\theta) =(1 - \delta\cos^2\theta)/(1 - \delta/2)$, 
where $\delta$ is the additional free parameter governing our 2D solutions that controls the magnitude of the density inhomogeneity.  
For this choice, we can evaluate the left hand side of eq. \eqref{rho_bound} after averaging this bound over $[0,\pi/2]$ to find 
$(2/\pi) \int_0^{\pi/2}|\pd{\ln f(\theta)}{\theta}| d\theta = (2/\pi)|\ln(1 - \delta)| \approx 2\delta/\pi$ (provided $\delta \ll 1$); eq. \eqref{rho_bound} becomes
\beq
 \delta \lesssim \f{2\pi}{R_o/R_B}  \left\langle \left| \f{v_r/c  + \pd{(v_\theta/c)}{\theta}}{T/T_0} \right| \right\rangle_{r = R_o},
\label{delta_bound}
\seq
where $\langle \cdot \rangle$ denotes an average from $[0,\pi/2]$.

We present results for five runs: 
three with a constant entropy BC (A2, A3, \& A4) and two with a constant pressure BC (B2 \& B3; the number denotes the density contrast, $-\log\delta$).
These BCs are implemented by assigning different pressures as a function of $\theta$ at $R_o$; we use $p(R_o,\theta) = p_0 [\rho_{\rm{bdy}}(\theta)/\rho_0]^\gamma$ for runs A2, A3, \& A4 and $p(R_o,\theta) =  p_0$ for runs B2 \& B3, where $p_0 = GM_{\rm{bh}}\rho_0/\gamma R_B$.   

\begin{figure*}
\includegraphics[width=\textwidth]{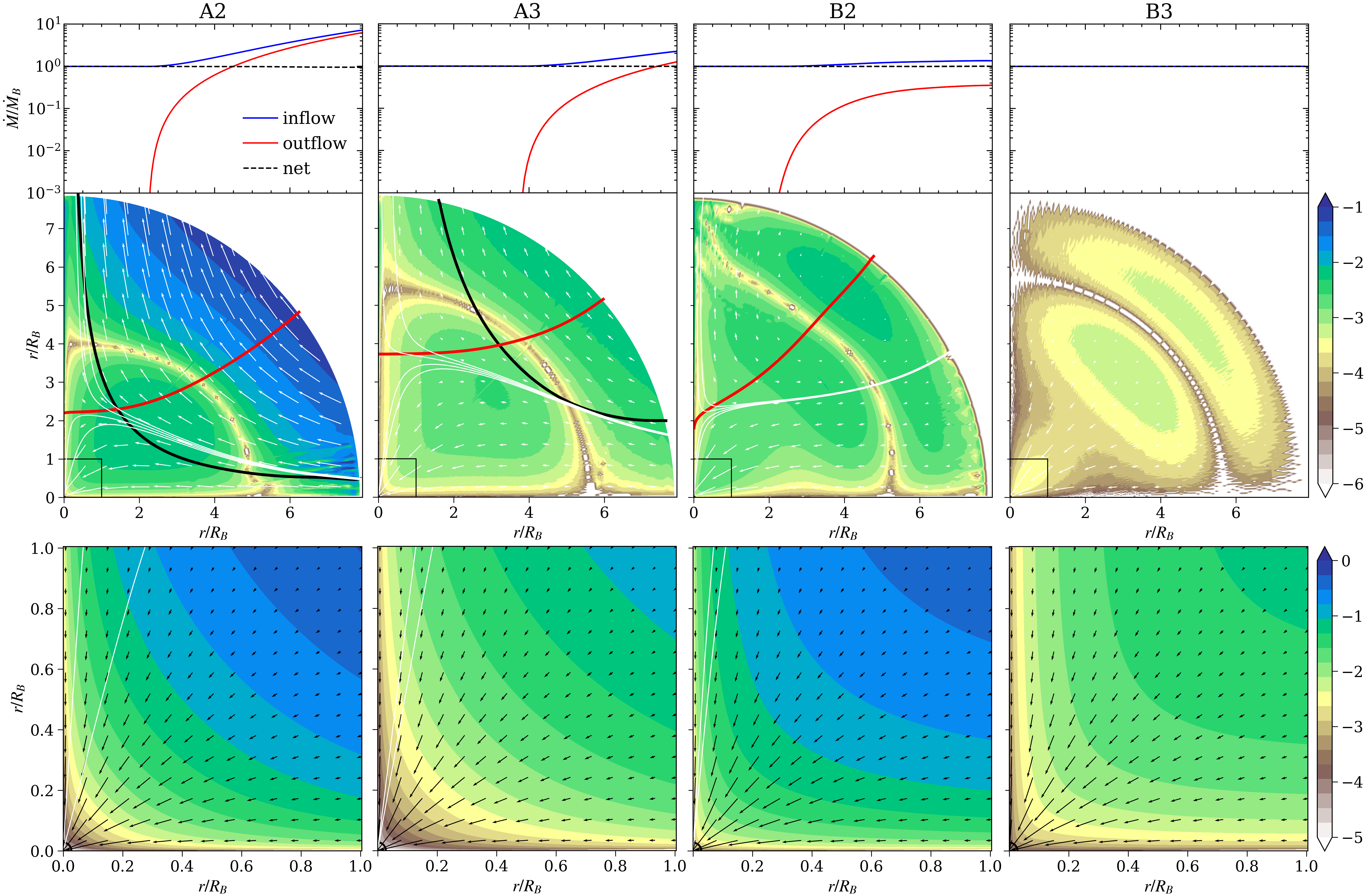}
\caption{
Steady state dynamics of our 2D solutions.  
Top panels: inflow, outflow, and net accretion rates as a function of radius.  
Middle panels: Log-scale colormaps of the ratio of the lateral pressure force and gravity, $|r^{-1}\pd{p}{\theta}/(GM\rho/r^2)|$, with the velocity field overplotted.
Red curves are contours of $v_r = 0$ and black curves are contours of $\ell_\phi = 2R_s c$.  The inner border along these curves approximates the true `Bondi surface' since regions interior to both curves have purely inflowing streamlines.  This is made clear by plotting closely spaced streamlines around the inflow/outflow turning point (white curves in runs A2, A3, \& B2).
Bottom panels: Log-scale colormaps of angular momentum ($\ell_\phi$) in units of $R_s c$ zoomed-in on the region within the Bondi radius ($R_B = 10^3\,R_s$) as marked by the rectangles above.  
The velocity field is again overplotted.  Note that $\ell_\phi$ is less than $R_s c$ and yet is a strongly varying function of $(r,\theta)$ despite other flow fields being closely spherically symmetric.
Animations of these runs are viewable at https://trwaters.github.io/OutflowsFromInflows/.
} 
\end{figure*}

\begin{table}
\centering
$\gamma = 5/3$, $R_B = 10^3R_s$, $R_o = 8 R_B$
\begin{tabular}{ccccccc}
\hline\\[-7pt]
Run & $\delta$ & $\delta/\delta_{\rm{c}}$ &
$\dot{M}_{\rm{in}}$  & $\dot{M}_{\rm{out}}$ & 
$\overline{M}_o$  
\\[4pt]\hline\hline\\[-7pt]
A2  & $10^{-2}$ & $7.8$ & 7.15 & 6.22 & 0.14   \\ [1pt]
A3  & $10^{-3}$ & $3.6$ & 2.22 & 1.26 & 0.032    \\ [1pt]
A4  & $10^{-4}$ & $1.7$  & 0.98 & $8.3\times 10^{-4}$ & $7.5\times 10^{-4}$   \\ [1pt]
\hline\\[-8pt]
B2 & $10^{-2}$ & - & 1.34 & 0.36 & 0.013  \\ [1pt]
B3 & $10^{-3} $ & - & 0.98 & 0.0  & -  \\ 
\hline
\end{tabular}
\caption{Properties of steady state solutions based on output at time $250 R_B/c_{s,0}$.  
Runs A2, A3, \& A4 (B2 \& B3) are for constant specific entropy (pressure) along $\theta$ at $R_o$.
The value of $\delta_{\rm{c}}$ is found by evaluating the right hand side of eq. \eqref{delta_bound}.  
The rates $\dot{M}_{\rm{in}}$ and $\dot{M}_{\rm{out}}$ are measured at $R_o$ in units of $\dot{M}_B$ and the accretion rate is $\dot{M}_{\rm{A}} = \dot{M}_{\rm{in}} - \dot{M}_{\rm{out}}$.  
Multiplying $\overline{M}_o$, 
the mean outflow Mach number at $R_o$ (where the overbar denotes averaging only over outflowing zones), by $(R_o/2R_B)^{1/2} = 2$ converts to velocity in units of the escape speed at $R_o$.  
The gas is unbound since the Bernoulli parameter at $R_o$ is $Be = 1.37c_{s,0}^2 > 0$ for all of these runs; $Be$ is only a weak function of radius, obeying $\pd{Be}{r} = \ell_\phi \omega_\phi/r$ (see eq. \eqref{radial_eqn}), where $\omega_\phi$ is the $\phi$-component of vorticity.  
}
\end{table}

\vspace{-0.1in}
\subsection{Solutions with constant entropy BCs}
Table~1 summarizes the properties of our five runs.   For solutions with homentropic BCs (runs A2, A3, \& A4), the bound given by eq. \eqref{delta_bound} is first formally satisfied for $\delta$ slightly below $10^{-4}$, meaning $\delta/\delta_c$ falls below unity, $\delta_c$ being the value of the right hand side of eq. \eqref{delta_bound}.   
Note that $\delta_c$ must be evaluated a posteriori since $v_r(R_o,\theta)$ and $\pd{v_\theta(R_o,\theta)}{\theta}$ are determined as part of a self-consistent solution.
Table~1 shows that run A4 has a very weak outflow, consistent with $\delta/\delta_c$ being somewhat larger than $1$.  

Steady state solutions for runs A2 and A3 are plotted in Fig.~1.  
The result of violating the angular momentum bound as measured by $\delta/\delta_c$ is an increasingly strong outflow.  
The top panels compare the inflow, outflow, and net accretion rates $(\dot{M}_{\rm{in}},\dot{M}_{\rm{out}},\dot{M}_{\rm{A}})$ as a function of radius.  
At $R_o$, run A3 has $\dot{M}_{\rm{out}} \approx 1.3\dot{M}_{\rm{B}}$, while A2 has $\dot{M}_{\rm{out}} \approx 6\dot{M}_{\rm{B}}$ (see the red lines).  
The net accretion rate (dashed black line) remains constant at nearly $\dot{M}_{\rm{B}}$, and this is possible in a steady state only if $\dot{M}_{\rm{in}} - \dot{M}_{\rm{out}} = \dot{M}_{\rm{A}} \approx \dot{M}_{\rm{B}}$ at every radius.    

The only force acting in the $\theta$-direction that can drive outflow by turning streamlines around is the gas pressure force $-r^{-1}\pd{p}{\theta}$.  The middle panels in Fig.~1 show maps of the ratio of this force to gravity.   Notice it is significant mainly in the outflowing regions at large radii.  It changes sign between $4-6R_B$, as it must transition from turning the flow outward to inward.   
The red and black contours denote where $v_r = 0$ and $\ell_\phi = 2 R_s c$, respectively.  Together these contours define an effective `Bondi surface': the region internal to both contours has purely inflowing streamlines, while streamlines are outflowing in the external regions.  This is seen clearly by identifying streamlines where the flow `turns over' (white curves); this turnover point closely follows the red and black contours as $\delta$ is changed.  

The bottom panels are maps of $\ell_\phi$ in units of $R_s c$, zoomed-in on the region within $1R_B$ where the accretion is nearly spherical.   These plots reveal that the flow is stratified in angular momentum, with $\ell_\phi$ increasing outward.  This dynamics is somewhat surprising, as gas at $r \lesssim R_B/2$ has $\ell_\phi \ll R_s c$, so intuitively angular momentum need not be transported outward for it to accrete.    
All that is required of these solutions from \S{2} is that they
satisfy $\int_S \rho\, \bm{\ell}\, \vv{v}\cdot d\vv{S} =0$, and since $\rho$ and $v_r$ are nearly spherically symmetric at $r < R_B$, this integral is approximately $2\pi \rho v_r r^2 \int_0^\pi \ell_\phi \sin\theta d\theta$.  The trivial way to obtain $\int_0^\pi \ell_\phi \sin\theta d\theta = 0$ for $\ell_\phi \neq 0$ is for $\ell_\phi$ to be an odd function in the range $[0,\pi]$, and this is automatic with an axis of symmetry at $\theta =\pi/2$.  

\vspace{-0.1in}
\subsection{Solutions with constant pressure BCs}
Constant pressure BCs are arguably more physical than constant entropy BCs, but in this case we can no longer derive a bound on $\pd{\ln\rho_{\rm{bdy}}}{\theta}$.  This is because eq. \eqref{lphi_phi} now gives
\beq
l_\phi = -R_o\f{\pd{h}{\theta} - T\pd{s}{\theta} + R_o\, v_r \, \pd{v_\theta}{r} }{v_r  + \pd{v_\theta}{\theta}} ,
\label{lphi_p}
\seq
and the first two terms in the numerator are simply $\rho^{-1}\pd{p}{\theta}$ by the basic thermodynamic relation $\rho^{-1}dp = dh - T\,ds$.  Thus, if we choose to assign a constant pressure at $R_o$, we are left with 
only an implicit relationship between $\ell_\phi$ and $\pd{\rho}{\theta}$ through $\pd{v_\theta}{\theta}$ via the continuity equation.  
Moreover, to allow for the possibility that $\ell_\phi \neq 0$ at $R_o$ when $\pd{p}{\theta} = 0$, we apply a constant gradient BC to $v_\theta$ instead of the zero gradient BC $(\pd{v_\theta}{r}=0)$ used in \S{3.1}.

Because a lateral gas pressure force ($=-R_o^{-1}\pd{p}{\theta}$) accompanies a constant entropy BC but not a constant pressure BC at $R_o$, we expect the outflow to be weaker for the same density BC.  This is indeed the case, as Table~1 and the right two sets of panels in Fig.~1 show.  Notice there is no black contour for runs B2 and B3 because all of the angular momentum in the domain is less than $2 R_s c$.  Nevertheless, as a consequence of the density gradient $\pd{\rho}{\theta}$ at $R_o$, a lateral pressure force still builds up enough to cause a weak outflow in run B2 (qualitatively similar to that of run A3).   
We verified that an outflow as strong as in run A2 under constant pressure BCs can be obtained by increasing $\delta$.

\section{Discussion}
The adiabatic solutions we find overall consist of an inner purely inflowing region with $\ell_\phi \lesssim R_s c$ that is engulfed in a much larger region containing all of the (still low angular momentum) flow with $\ell_\phi > R_s c$.     
Rather than circularize, as would happen in solutions where the angular momentum is due to $v_\phi$,    
the flow with $\ell_\phi > R_s c$ forms an inflow-outflow region outside an effective `Bondi surface', this being how angular momentum is transported outward.  In recent papers following up an investigation by Hernandez et al. (2014) using imposed density gradients as a mechanism for launching hydrodynamic jets, Aguayo-Ortiz et al. (2019) and Tejeda et al. (2019) found an incompressible analytic solution that captures the inflow-outflow topology of adiabatic solutions.  However, these authors do not discuss the role of angular momentum, which is essential to understanding our results.  

Our simulations show that the accretion rate onto the BH remains constant at nearly $\dot{M}_B$, in agreement with the findings of Tejeda et al. (2019), although we do not expect this result to hold for simulations with a nonzero $v_\phi$ at $R_o$.  We therefore caution against concluding that the Bondi formula provides an adequate estimate for the accretion rate onto `BH particles' in cosmological simulations.  More relevant for the construction of improved sub-grid models of BH growth is our result that the outflow rate will always exceed $\dot{M}_B$ for realistic density contrasts; this finding is in agreement with the 3D simulations of stellar wind fed accretion onto Sgr A* by Ressler et al. (2019).  
We note that SED-fitting results for Sgr A* favor models in which the outflow transitions from weak to strong as radius increases (Ma et al. 2019), consistent with the dynamics we find here.  

For a mean outflow velocity given by $\overline{M}_o c_{s,0}$ (with $\overline{M}_o$ the mean Mach number of the outflow at $R_o$), the outflow rate should scale as $\dot{M}_{\rm{out}} = \dot{M}_B  \overline{M}_o (R_o/R_B)^2$. 
In addition to the runs presented, which have $R_o/R_B = 8$, we examined variants of run A2 with $R_o/R_B$ halved and doubled to verify this scaling.  It is important to confirm that this scaling holds in Bondi-Hoyle type simulations with an imposed wind
(the properties of which determine the conditions along $R_o$), as well as to determine if the associated feedback can in turn limit the accretion rate by affecting the mass reservoir.  
We note that for supersonic motion through an inhomogenous cloud, as considered in a few studies of 3D Bondi-Hoyle accretion (e.g., Ruffert 1999; Xu \& Stone 2019), the imposed ram pressure may be too high and/or the density gradients too low to allow inflow-outflow regions to develop.\footnote{Specifically, taking $R_o = GM_{\rm{bh}}/v_{\rm{\infty}}^2$ (the Hoyle-Littleton radius for relative motion through a cloud at speed $v_{\rm{\infty}}$), the quantity $(R_o/R_B)^2 = 1/M_{\rm{in}}^4$, where $M_{\rm{in}} = v_{\rm{\infty}}/c_{s,0}$ is the characteristic Mach number of the mass reservoir, suggesting that inflow-outflow regions with $\dot{M}_{\rm{out}} > \dot{M}_B$ will develop only for $M_{\rm{in}} \lesssim 1$.}  
For marginally supersonic motion, however, this outflow dynamics should occur, and 
it is unclear what effect it will have on the instabilities accompanying Bondi-Hoyle accretion.

\vspace{-0.2in}
\section*{Acknowledgments}
TW thanks Ya-Ping Li and Josh Dolence for helpful discussions.
TW and AA are partially supported by 
the LANL LDRD Exploratory Research Grant 20170317ER.    
DP acknowledges support for program number HST-AR-14579.001-A provided by NASA through a grant from the Space Telescope Science Institute, which is operated by the Association of Universities for Research in Astronomy, Incorporated, under NASA contract NAS5-26555.
HL is supported by NASA/ATP and a LANL LDRD.  

\vspace{-0.2in}
\appendix
\section{Radial and transverse flow equations}
For simulations performed in spherical coordinates, it is useful to derive equations governing the flow on and normal to any spherical surface in a steady state, which follow from Crocco's theorem (see e.g., Rezzolla \& Zanotti 2013), 
\beq
\nabla Be =  \vv{v} \times \vor  + T\nabla s +  \vv{f}_b,
\label{Crocco_thm}
\seq
where $Be \equiv h + v^2/2 + \Phi$ is the Bernoulli function, $\vor \equiv \nabla \times \vv{v}$ the vorticity, and $\vv{f}_b$ the sum of both magnetic and radiation forces.  
Note that Crocco's theorem governs both adiabatic and non-adiabatic solutions and that the generalized Bernoulli's theorem follows
by dotting eq. \eqref{Crocco_thm} with $\vv{v}$ to give $\nabla Be =T\nabla s + \vv{f}_b$ along streamlines. 
By instead dotting eq. \eqref{Crocco_thm} with $\vv{r} = r\hat{r}$, we obtain the radial flow equation
\beq
\vv{r} \cdot \nabla Be = \bm{\ell} \cdot \vor + T\vv{r}\cdot\nabla s +  \vv{r}\cdot\vv{f}_b.
\label{radial_eqn}
\seq
In the Bondi solution, all terms on the right hand side are zero and $\vv{v}\parallel \vv{r}$, so this reduces to the classical Bernoulli theorem, namely $Be = constant$ along streamlines.

Taking the cross product of $\vv{r}$ and eq. \eqref{Crocco_thm} gives
\beq
\vv{r}\times \nabla \left( h + \f{v^2}{2} \right) =  (\vv{v}\cdot \vor)\vv{r} - r\,v_r\vor - \vor \times \bm{\ell}  +  T\vv{r}\times\nabla s +  \bm{\tau},
\label{transverse_eqn}
\seq
where the identities $\vv{r} \times (\vv{v} \times \vor) = \vv{v}\times(\vv{r}\times\vor) - \vor \times \bm{\ell} $ and $\vv{v}\times(\vv{r}\times\vor) = (\vv{v}\cdot \vor)\vv{r} - (\vv{v}\cdot\vv{r})\vor$ were used.  Here, $\bm{\tau} = \vv{r}\times\vv{f}_b$ is the torque from any magnetic or radiation forces.  
Note that the radial component of this equation is an identity.  
In axisymmetry, the $\theta$-component is  
$\tau_\theta/r = v_r \omega_\theta - v_\theta\omega_r$, and both $\omega_r$ and $\omega_\theta$ are zero when $v_\phi = 0$, leading to an inconsistency if $\tau_\theta \neq 0$.  This is the intuitive statement that steady state solutions with $v_\phi = 0$ are impossible if there is any net applied torque in the poloidal plane, for meridional circulation will just continually build up.  
From the $\phi$-component,
we can arrive at an expression for $\ell_\phi = r v_\theta$ for axisymmetric flows that holds on every spherical surface,
\beq
l_\phi = r\f{v_\phi^2\cos\theta - r v_r \sin\theta\, \pd{v_\theta}{r} + S(\vv{r})\sin\theta}{(v_r  + \pd{v_\theta}{\theta})\sin\theta},
\label{lphi_phi}
\seq
where $S(\vv{r}) =  T\pd{s}{\theta} - \pd{h}{\theta}  + \tau_\phi$.   
Notice from eq. \eqref{lphi_phi} that stagnation points (where $\vv{v} = 0$) can only occur at locations $\vv{r}_s$ satisfying $S(\vv{r}_s) = 0$, meaning there is a lateral force balance, $\pd{p}{\theta} = \rho \tau_\phi$, while there is simultaneously a radial force balance by eq. \eqref{radial_eqn}.

\end{document}